\documentclass[12pt,a4paper,final]{iopart}
\usepackage{iopams}
\usepackage{graphicx}
\usepackage[breaklinks=true,colorlinks=true,linkcolor=blue,urlcolor=blue,citecolor=blue]{hyperref}

\begin{document}

\title[]{Interplay between nematic fluctuation and superconductivity in the two-orbital Hubbard
model: A quantum Monte Carlo study}

\author{Guangkun~Liu$^{1}$, Shichao~Fang$^{2}$, Xiaojun~Zheng$^3$,
Zhongbing~Huang$^{1,2}$, Haiqing~Lin$^1$}
\address{$^1$Beijing Computational Science Research Center, Beijing 100193, China}
\address{$^2$Faculty of Physics and Electronic Technology, Hubei University, Wuhan 430062, China}
\address{$^3$Guilin University of Technology, Guilin 54100, China}

\ead{huangzb@hubu.edu.cn}

\begin{abstract}
To understand the interplay between nematic fluctuation and superconductivity in iron-based superconductors,
we performed a systematic study of the realistic two-orbital Hubbard model by using the constrained-path
quantum Monte Carlo method. Our numerical results showed that the on-site nematic interaction induces a
strong enhancement of nematic fluctuations at various momentums, especially at ($\pi$,$\pi$).
Simultaneously, it was found that the on-site nematic interaction suppresses the ($\pi$,0)/(0,$\pi$)
antiferromagnetic order and long-range electron pairing correlations for dominant pairing channels in
iron-based superconductors. Our findings suggest that nematic fluctuation seems to compete with
superconductivity in iron-based superconductors.
\end{abstract}


\section{Introduction}
Iron-based superconductors (FeSCs) continue to attract the interests of condensed
matter community~\cite{Dagotto2013,Johnston2010,Hirschfeld2011,Hirschfeld2011,Dai2012}.
One common strategy to understand the superconducting phase in FeSCs is to study
the normal states where superconductivity arises. For most FeSCs, superconductivity
is found in proximity to a nematic state, in which the systems spontaneously break
the rotational symmetry and preserve time-reversal symmetry below certain
temperatures~\cite{Lawler2010,Fischer2011,Edegger2006,Sun2010,Zheng2014,Zheng2015}.
Debating about the origin of nematicity still exists among spin-nematic
~\cite{Fernandes2011,Fernandes2014,Fernandes2012} ,ferro-orbital order
~\cite{Lee2012,Dai2012,Lv2009,Lv2010} and other scenarios~\cite{Bishop2016,Bishop2017}.
Many experimental evidences indicate that nematicity and superconductivity
have a common microscopic origin. For instance, angular-dependent magnetoresistance
and static magnetization measurement on FeSe samples showed that the onset
temperature T$_{n}$ of nematic order has a universal linear relationship with
the superconducting transition temperature T$_c$~\cite{Yuan2016}. Therefore, it
is essential to understand the nematic state as it may play an important role
to understand superconductivity.

Regarding the relationship between nematicity and superconductivity, it is
still under debate~\cite{Lederer2015, Kaczmarczyk2016}. Some experimental
and theoretical researches seem to support the coexisting scenario
between nematicity and superconductivity. For instance, McQueen et
al.~\cite{McQueen2009} reported the low temperature structural properties of
FeSe by high resolution synchrotron x-ray power diffraction,
transmission electron microscopy, and electron diffraction. Their results
indicated a coexistence of superconductivity and nematic order. Lederer et
al.~\cite{Lederer2015} considered a low T$_c$ metallic superconductor weakly
coupled to the soft fluctuations associated with proximity to a
nematic quantum critical point (NQCP) and found an enhancement of
superconductivity near the NQCP. On the other hand, many researchers also
found evidences for the competition between superconductivity and
nematicity. For example, Kim et al.~\cite{Kim2014} studied the evolution of
the temperature dependence of the in-plane London penetration depth,
$\Delta\lambda(T)$ in high-quality single crystals of Ba$_{1}$K$_x$Fe$_2$As$_2$
and found a power law behavior of $\Delta\lambda(T)$ in the under doped region,
indicating a competition between nematicity and superconductivity. Besides,
Cai et al.~\cite{Cai2014} studied the doping dependence of quasiparticle
interference (QPI) in NaFe$_{1-x}$Co$_x$ and the QPI pattern at optimal
doping is still fourfold symmetric, which suggests that nematic fluctuation
is not a prerequisite for electron pairing. Moon et al. ~\cite{Moon2012}
presented a general theory of competition between superconductivity
and nematic order, in which the concomitant instabilities of both orders are
produced by the underlying Fermi surface.

In this work, we will not focus on the origin of nematic state, instead,
by introducing nematic fluctuation to the realistic two-orbital Hubbard model through
an on-site nematic interaction, we are trying to explore the important issue
concerning the relationship between nematicity and superconductivity. Our motivation
comes from recent quantum Monte Carlo (QMC) and random phase approximation (RPA)
studies on a simple two-orbital model~\cite{Dumitrescu2016,Yamase2013}, which only
considered the electron hopping and on-site nematic interaction terms. The model offers
a new way to analyze the effect of nematicity, however, the coupling of electron correlation
and nematicity in FeSCs calls for inclusion of electron Coulombic interactions in the
microscopic model. Our results based on the constrained-path quantum Monte Carlo method
(CPQMC)~\cite{Zhang1997a,Zhang1997b} show that the introduced on-site nematic interaction
would induce a strong enhancement of orbital fluctuations. Such fluctuations decrease the
($\pi$,0)/(0,$\pi$) magnetic order and also suppress the possible long-range electron
pairings. Our findings suggest that there exists a competition between nematic fluctuation
and superconductivity in the studied models.

The organization of this paper is as follows: the introduced two-orbital
Hubbard model is defined in Section~I.  The technical details of CPQMC method is
described in Section~III. Section~IV contains our numerical results, and finally in
Section~V, we provide further discussions and present our conclusions.

\section{Model and Method}
We will focus on the two-orbital Hubbard model for FeSCs, together with an
on-site nematic interaction which was introduced in Refs.~\cite{Yamase2013,Dumitrescu2016}.
Briefly, the model is composed of the tight-binding $H_{\mathrm{t}}$, the on-site
Coulombic interactions $H_{\mathrm{Coul}}$, and the on-site nematic interaction
$H_{\mathrm{nem}}$. The full Hamiltonian is expressed as
$H=H_{\mathrm{t}}+H_{\mathrm{Coul}}+H_{\mathrm{nem}}$.

The tight-binding component is described as
\begin{eqnarray}
H_{\mathrm{t}}=&
-t_1\sum_{\textbf{i},\sigma}
(d_{\textbf{i},xz,\sigma}^{\dagger}d_{\textbf{i}+\hat{y},xz,\sigma}+
 d_{\textbf{i},yz,\sigma}^{\dagger}d_{\textbf{i}+\hat{x},yz,\sigma}+\mathrm{h.c.})
\nonumber\\
&-t_2\sum_{\textbf{i},\sigma}
(d_{\textbf{i},xz,\sigma}^{\dagger}d_{\textbf{i}+\hat{x},xz,\sigma}+
 d_{\textbf{i},yz,\sigma}^{\dagger}d_{\textbf{i}+\hat{y},yz,\sigma}+\mathrm{h.c.})
\nonumber\\
&-t_3\sum_{\textbf{i},\hat{\mu},\hat{\nu},\sigma}
(d_{\textbf{i},xz,\sigma}^{\dagger}d_{\textbf{i}+\hat{\mu}+\hat{\nu},xz,\sigma}+
 d_{\textbf{i},yz,\sigma}^{\dagger}d_{\textbf{i}+\hat{\mu}+\hat{\nu},yz,\sigma}+
 \mathrm{h.c.}) \nonumber\\
&+t_4\sum_{\textbf{i},\sigma}
(d_{\textbf{i},xz,\sigma}^{\dagger}d_{\textbf{i}+\hat{x}+\hat{y},yz,\sigma}+
 d_{\textbf{i},yz,\sigma}^{\dagger}d_{\textbf{i}+\hat{x}+\hat{y},xz,\sigma}+
 \mathrm{h.c.}) \nonumber\\
&-t_4\sum_{\textbf{i},\sigma}
(d_{\textbf{i},xz,\sigma}^{\dagger}d_{\textbf{i}+\hat{x}-\hat{y},yz,\sigma}+
 d_{\textbf{i},yz,\sigma}^{\dagger}d_{\textbf{i}+\hat{x}-\hat{y},xz,\sigma}+
 \mathrm{h.c.}),
\end{eqnarray}
where $xz$ and $yz$ denote the $d_{xz}$ and $d_{yz}$ orbitals, respectively.
The operator $d_{\textbf{i}\alpha\sigma}^\dagger$ ($d_{\textbf{i}\alpha\sigma}$)
creates (annihilates) an electron on orbital $\alpha$ in Fe site $\textbf{i}$
with spin $\sigma$, and the index $\hat{\mu}(\hat{\nu})=\hat{x}$ or $\hat{y}$
denotes a unit vector linking the nearest-neighbor sites. In order to
gain a full understanding of FeSCs, we adopt two sets of hopping
parameters~\cite{Raghu2008,Dumitrescu2016}: one is taken as
$t_1=-1.0$, $t_2=1.3$ and $t_3=t_4=-0.85$, which we marked as Raghu
hopping parameters, and the other is set as $t_1=-1.0$, $t_2=1.5$, $t_3=-1.2$,
and $t_4=-0.95$, which we marked as Dumitrescu hopping parameters.

The Coulombic interaction $H_{\mathrm{Coul}}$ is defined as

\begin{eqnarray}\label{hint2}
H_{\mathrm{Coul}} = &
U\sum_{\textbf{i}\alpha}
n_{\textbf{i}\alpha\uparrow}n_{\textbf{i}\alpha\downarrow}
+\left(U'-J/2\right)\sum_{\textbf{i}}
n_{\textbf{i},xz}n_{\textbf{i},yz}
-2J\sum_{\textbf{i}}
\mathrm{S}_{\textbf{i},xz}^{\mathrm{z}}
\mathrm{S}_{\textbf{i},yz}^{\mathrm{z}},
\end{eqnarray}
where $n_{\textbf{i}\alpha}=n_{\textbf{i}\alpha\uparrow}+
n_{\textbf{i}\alpha\downarrow}$ is the electron density operator
at orbital $\alpha$ ($\alpha = xz, yz$) on site $\textbf{i}$. The
$z$-component of spin operator is defined as
$\mathrm{S}_{\textbf{i},\alpha}^{\mathrm{z}}=\frac{1}{2}
(n_{\textbf{i}\alpha\uparrow}-n_{\textbf{i}\alpha\downarrow})$.
We also keep $U'=U-2J$ and $J=U/4$ as in previous literatures
~\cite{Luo2010,Liu2016,Daghofer2008}. Note that we simplified Hund's
coupling term, $\sum_{\textbf{i}} \mathrm{S}_{\textbf{i},xz}\mathrm{S}_{\textbf{i},yz}$,
to its Ising contribution, and also ignored the pair-hopping items.
This simplification are based on two observations~\cite{Liu2016,Liu2014}:
(1) previous QMC studies have shown that the Ising contribution of the Hund's
interaction could capture the main physics and (2) QMC simulations could
produce higher numerical accuracy.

Finally, in order to study the effect of nematic correlation,  the
on-site nematic interaction $H_{\mathrm{nem}}$ is added to the Hamiltonian,
which is defined as~\cite{Yamase2013,Dumitrescu2016}
\begin{eqnarray}
H_{\mathrm{nem}}=&-\frac{g}{2}\sum_{\textbf{i}}(n_{i,xz}-n_{i,yz})^2.
\end{eqnarray}
$H_{\mathrm{rem}}$ breaks the orbital symmetry and directly induces nematic
correlation without any prior orbital order.

For the magnetic and nematic properties, we examine the spin structure factor
and nematic structure factor as follows,
\begin{eqnarray}
S(\textbf{q})&=
1/N\sum_{ij} e^{i\textbf{q}\cdot(\textbf{r}_i-\textbf{r}_j)}
\langle(n_{\textbf{i}\uparrow}-n_{\textbf{i}\downarrow})
       (n_{\textbf{j}\uparrow}-n_{\textbf{j}\downarrow})\rangle,\\
N(\textbf{q})&=1/N\sum_{ij} e^{i\textbf{q}\cdot(\textbf{r}_i-\textbf{r}_j)}
\langle(n_{\textbf{i},xz}-n_{\textbf{i},yz})
       (n_{\textbf{j},xz}-n_{\textbf{j},yz})\rangle,
\end{eqnarray}\label{factors}
where $\textbf{q}$ and $\textbf{r}$ are the momentum and coordinate points,
respectively. $N$ counts the number of $\textbf{r}_i$ and $\textbf{r}_j$ pairs.

For the superconducting property, the classification of possible nearest neighbor
pairing symmetries in Ref.~\cite{Wan2009} is followed. The pairing operator
can be defined as ~\cite{Liu2014,Moreo2009}
\begin{equation}
\Delta^{\dagger }(\textbf{q})=
\frac{1}{\sqrt{2}}f(\textbf{q})(\tau_i)_{\alpha,\beta}
(d_{\textbf{q},\alpha,\uparrow}^{\dagger}d_{-\textbf{k},\beta,\downarrow}^{\dagger}-
 d_{\textbf{q},\alpha,\downarrow}^{\dagger}d_{-\textbf{k},\beta,\uparrow}^{\dagger}),
\end{equation}
where $d_{{\textbf{q},\alpha,\sigma}}^{\dagger}$ creates an electron
in orbital $\alpha$ with momentum $\textbf{q}$ and spin $\sigma$,
and $f(\textbf{q})$ is the form factor and $\tau_i$'s are the Pauli matrices
($i=1,2,3$) or identity matrix ($i=0$).

Using the Fourier transformation, we can get the
pairing operator in coordinate space $\Delta(\textbf{i})$, and the
corresponding pairing correlation function is defined as
$P(r=\left|\textbf{i}-\textbf{j}\right|)=\langle
\Delta^{\dagger}(\textbf{i})\Delta(\textbf{j})\rangle$.
We also calculated averaged pairing correlations through all
distances $P_{all}$ and long-range distances $P_{long}$ as
$P_{all}=\frac{1}{M}\sum_rP(r)$ and $P_{long}=\frac{1}{M'}\sum_{r>2}P(r)$,
with M and M$^{\prime}$ representing the numbers of $P(r)$.

\begin{table}
\caption{Definitions of four types of pairings considered in this paper.
The numbering of the pairing is simply adopted from Ref.~\cite{Wan2009}}
\centering
\begin{tabular}[b]{cccc}
\hline\hline 
No.&        IR&   $f(\textbf{k})\tau_i$&    \\[0.5ex]\hline
$s_{\pm}$&  $A_{1g}$&    $\cos k_x \cos k_y\tau_0$&       \\
    wave2&  $A_{1g}$&  $(\cos k_x+\cos k_y)\tau_0$&       \\
    wave3&  $A_{1g}$&  $(\cos k_x-\cos k_y)\tau_3$&       \\
    wave6&  $B_{1g}$&  $(\cos k_x-\cos k_y)\tau_0$&       \\[1ex]
\hline\hline
\end{tabular}
\label{pairnumber}
\end{table}

We study the Hamiltonian by using the CPQMC method, which is a sign-problem-free
auxiliary-field quantum Monte Carlo method. It projects out the ground state
from a trial state by branching random walks in the Slater determinant space.
A constrained-path approximation is adopted in the CPQMC algorithm to prevent the sign
problem~\cite{Zhang1997a,Zhang1997b}. For its application to multi-orbital Hubbard
models, we refer the readers to Refs.~\cite{Liu2016,Liu2014,Sakai2004}.
In a typical large-scale CPQMC simulation, we set the average
number of random walkers to be 4800 and the time step
$\Delta\tau=0.04$. 2000 Monte Carlo steps were sampled before
measurements, and 10 blocks of 480 Monte Carlo steps each were
used to ensure statistical independence during the measurements. Closed-shell
fillings were chosen in the simulations. To judge the accuracy of the CPQMC method,
we compared the CPQMC energies against those employing the Lanczos method
on a small systems: the maximum energy difference is within $1\%$ up to
$U=4.0$ eV.

\begin{figure}
\centering
\includegraphics[scale=0.62]{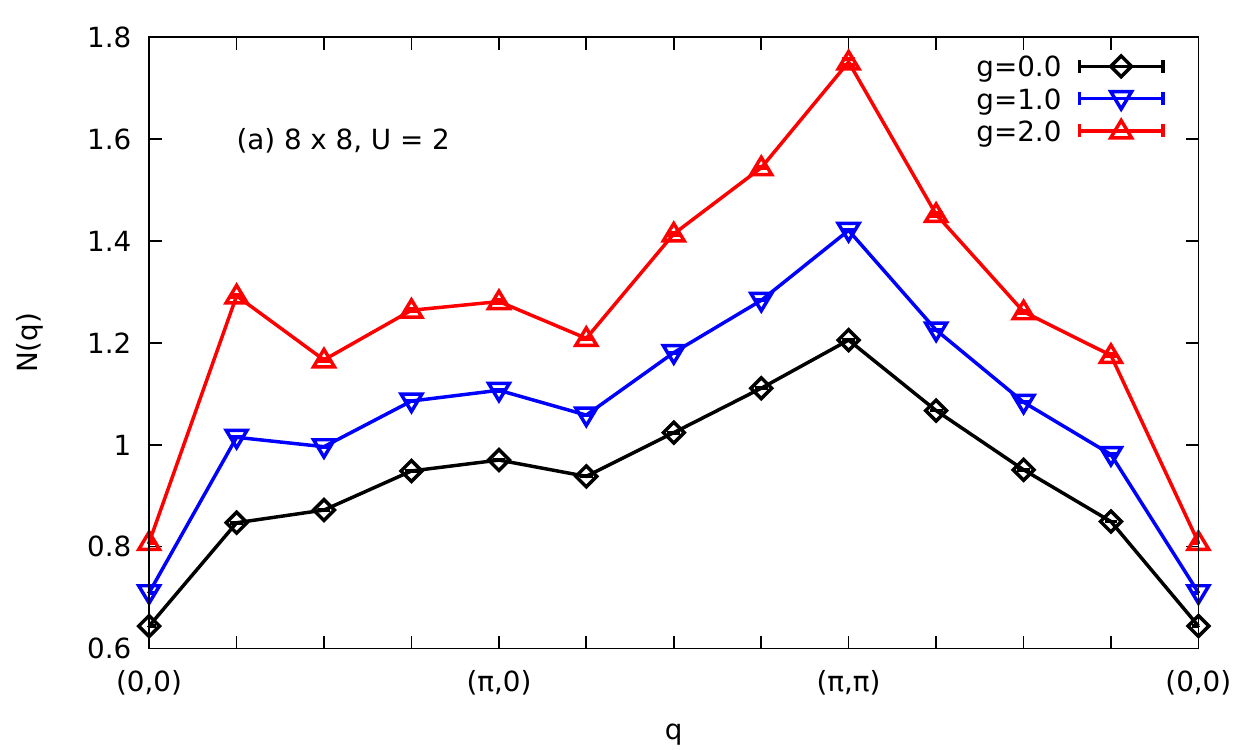}
\includegraphics[scale=0.62]{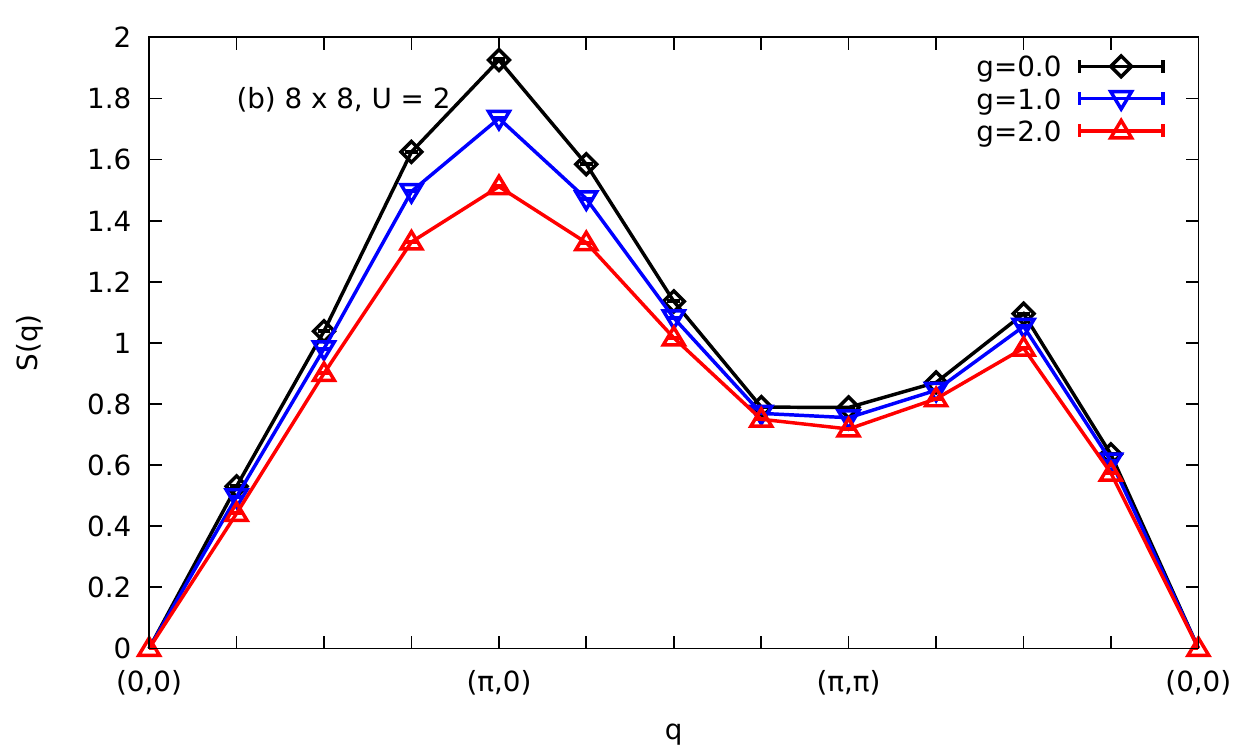}
\caption{(color online)
(a)Nematic structure factor N(q) and (b) Spin structure factor S(q) versus high
symmetric q-space points along the direction (0,0)-($\pi$,0)-($\pi$,$\pi$)-(0,0),
with hole doped rate $\rho=0.125$ and U $=2$ eV on an 8$\times$8 lattice.
Different symbols represent the values under different on-site nematic
correlation strength g. Periodic boundary conditions and close-shell filling
are used during the simulations. The results are obtained by using Dumitrescu
hopping parameters.} \label{struc_factor1}
\end{figure}

\begin{figure}
\centering
\includegraphics[scale=0.62]{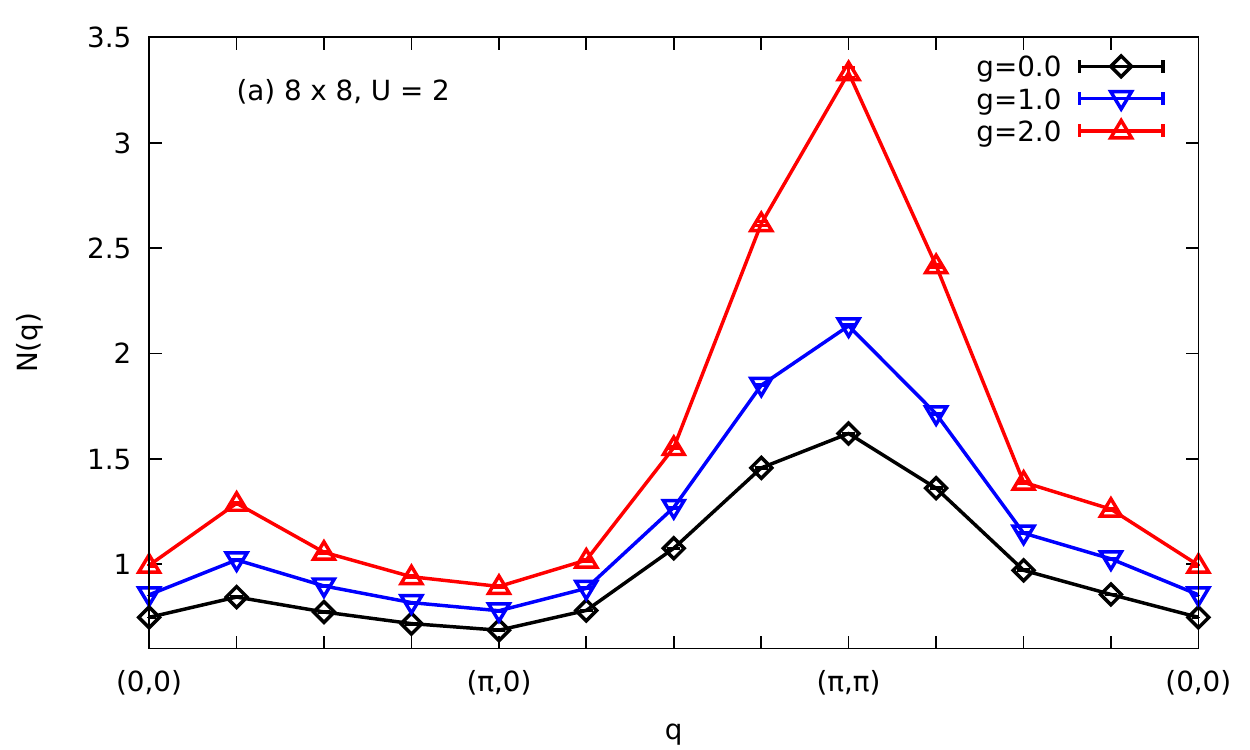}
\includegraphics[scale=0.62]{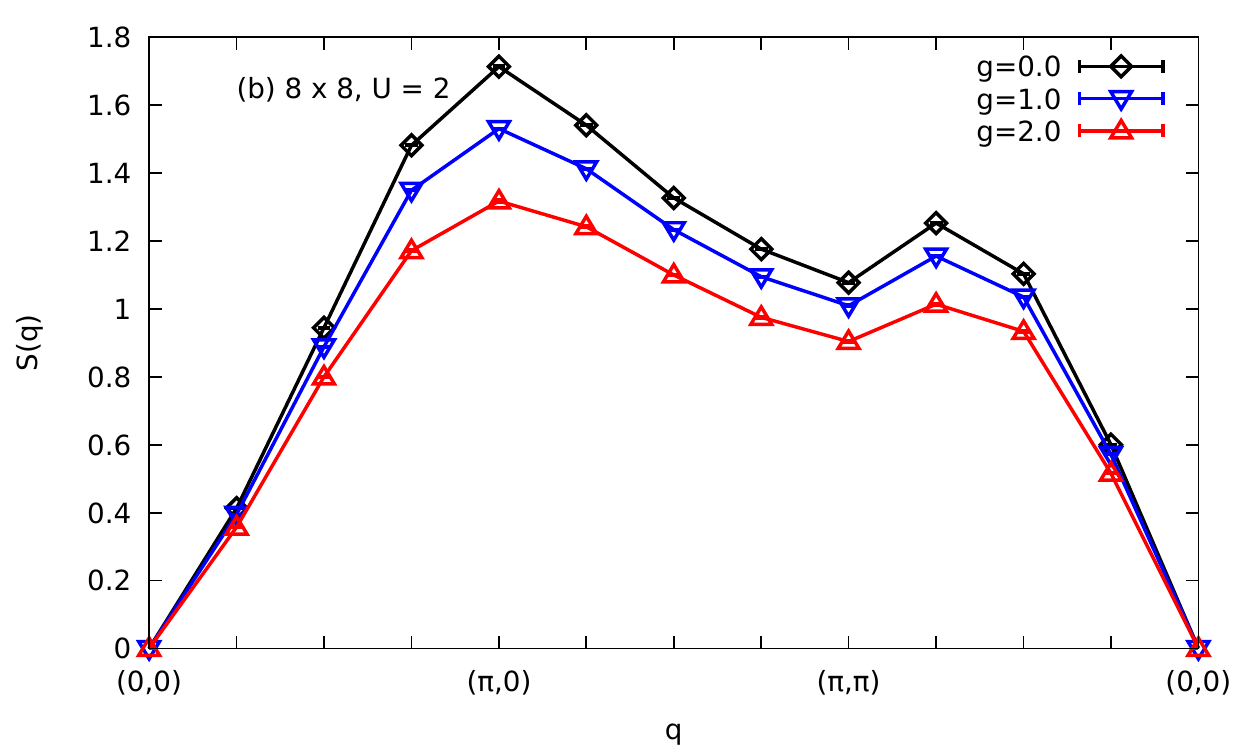}
\caption{(color online)
(a)Nematic structure factor N(q) and (b) Spin structure factor S(q) versus high
symmetric q-space points with the same doping density, lattice size, and on-site
electronic correlations as Fig.~\ref{struc_factor1}. The results are obtained by using
Raghu hopping parameters.
} \label{struc_factor2}
\end{figure}

\begin{figure}
\centering
\includegraphics[scale=0.62]{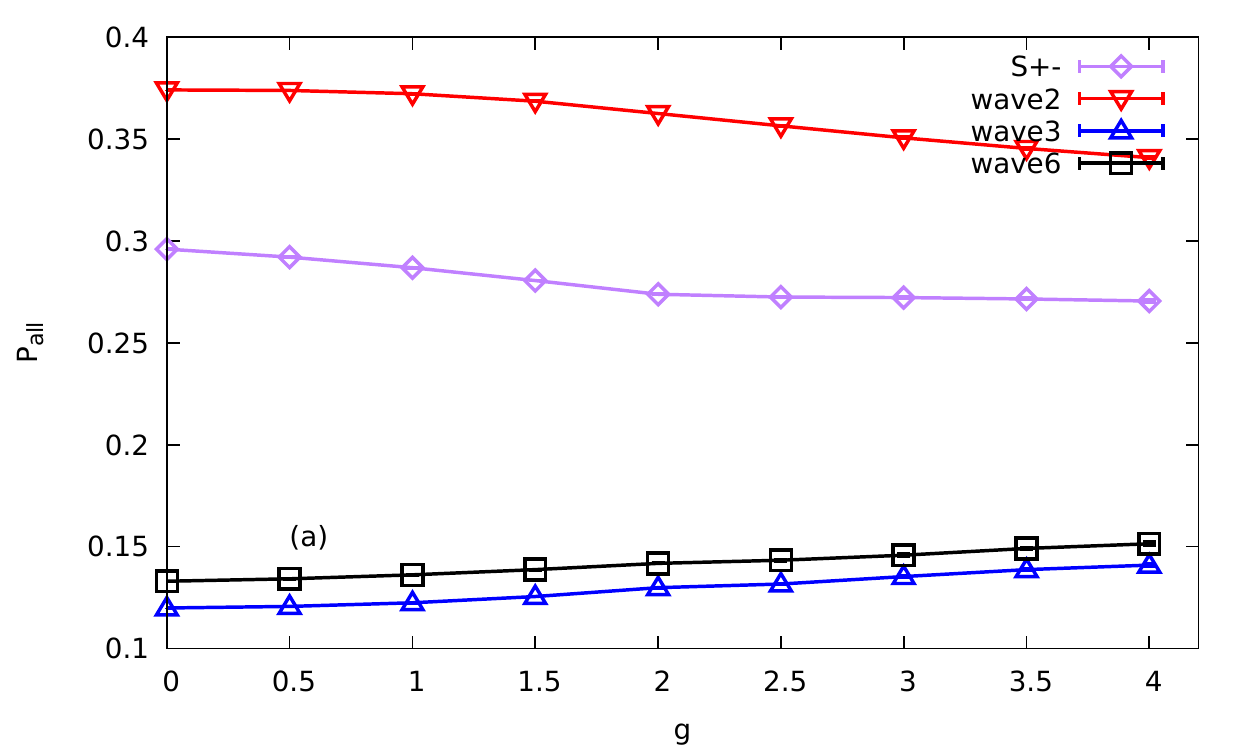}
\includegraphics[scale=0.62]{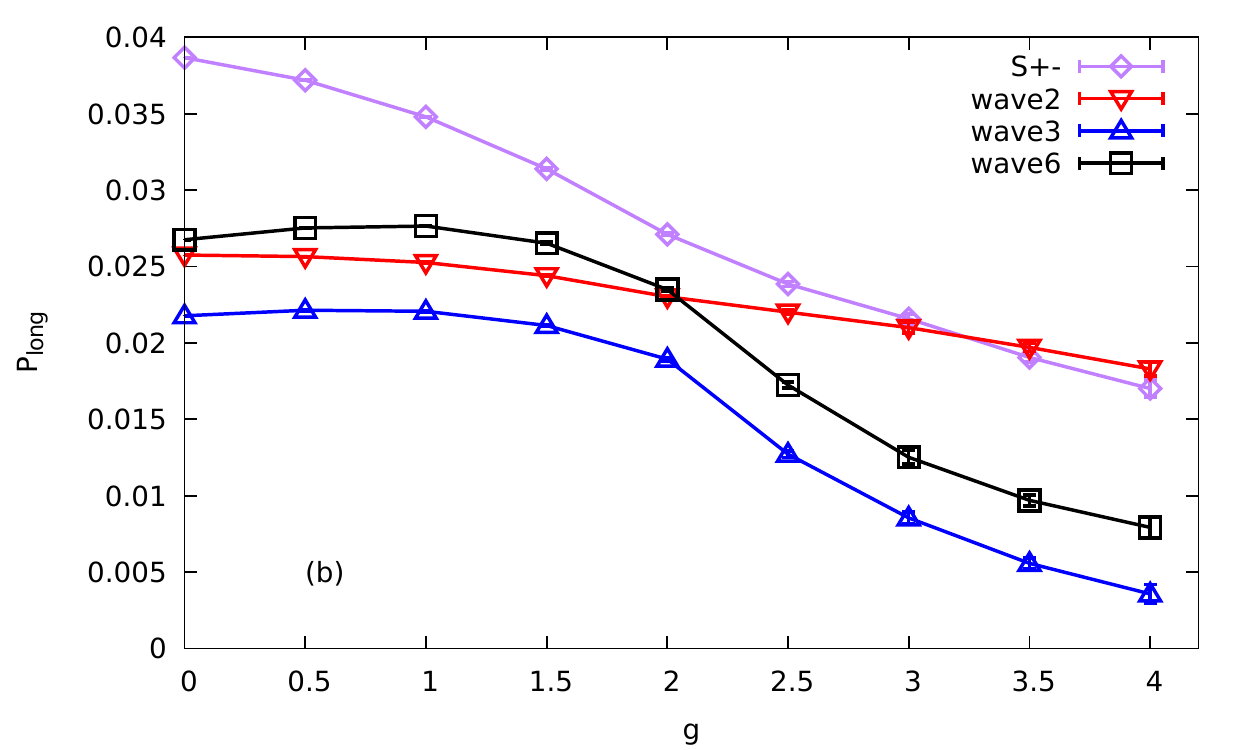}
\caption{(color online)
Averaged pairing correlation function through all pairing distances,
$P_{all}$, and the long-range-averaged pairing correlations, $P_{long}$,
for selected pairing channels with hole doping density $\rho=0.125$
and U = 2 eV on an 8$\times$8 lattice. Different symbols represent different
pairing channels, and the detailed definitions of the pairings  are given in
Table I and Eq. (5). Periodic boundary conditions and close-shell filling
are used during the simulations. The results are obtained by using Dumitrescu
hopping parameters.} \label{ave_p1}
\end{figure}

\begin{figure}
\centering
\includegraphics[scale=0.62]{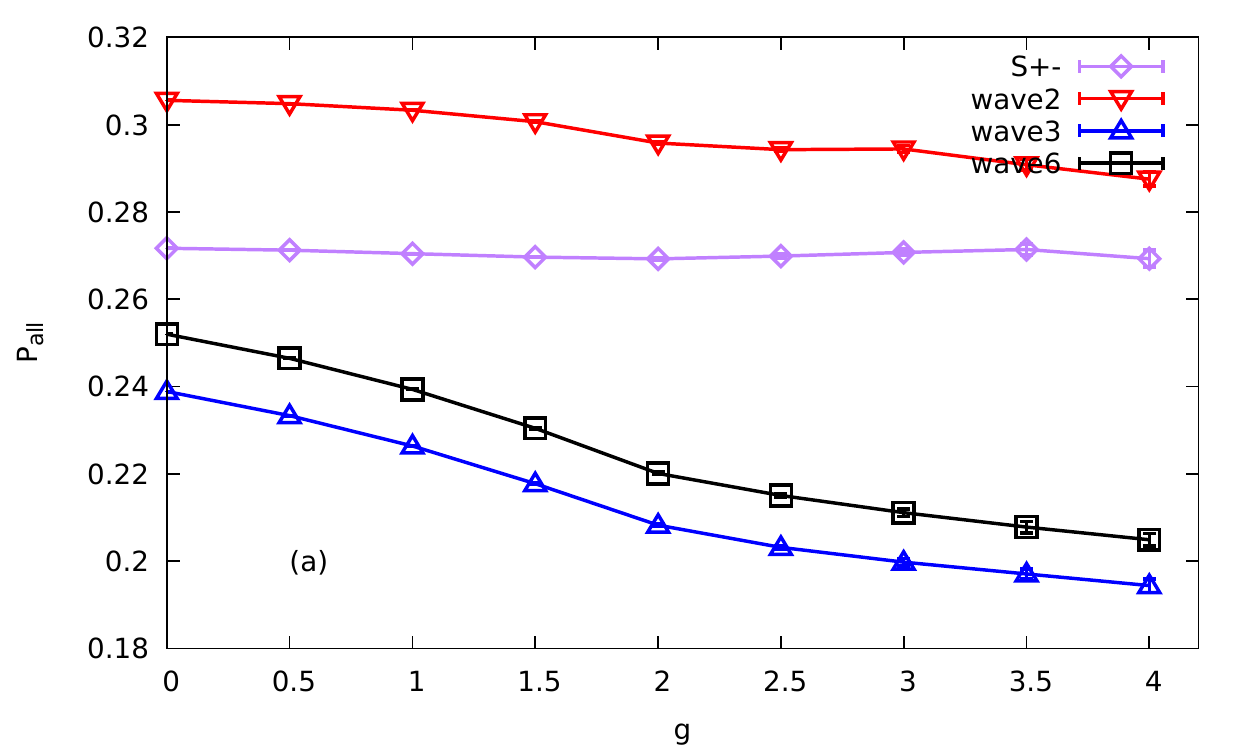}
\includegraphics[scale=0.62]{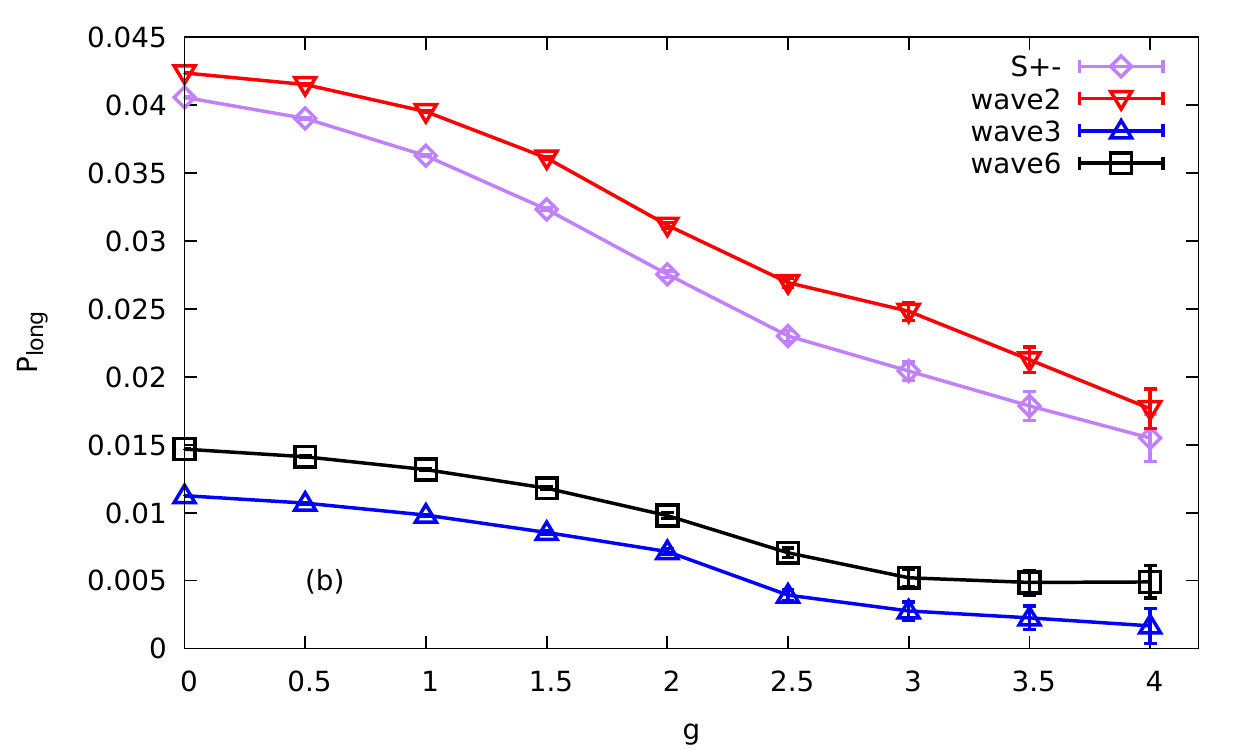}
\caption{(color online)
Averaged pairing correlation function through all pairing distances,
$P_{all}$, and the long-range-averaged pairing correlations, $P_{long}$, for
selected pairing channels with hole doping density $\rho=0.125$ and U = 2 eV on an
8$\times$8 lattice. Different symbols represent different pairing channels,
and the detailed definitions of the pairings see Table I and Eq. (5).
Periodic boundary conditions and close-shell filling are used during
the simulations. The results are obtained by using
Raghu hopping parameters.} \label{ave_p2}
\end{figure}

\section{Results}
\subsection{Nematic and spin correlations}
Firstly we check the effect of on-site nematic correlation in the doped
systems. By using Dumitrescu hopping parameters~\cite{Dumitrescu2016},
Fig.~\ref{struc_factor1}(a) illustrates the nematic structure factor along
the high symmetric momentum-space points. Increasing the on-site nematic
correlation g enhances all the nematic structure factors, especially for the
($\pi$,$\pi$) point. We also examined the real space nematic correlation versus
distance r as
$N(r=|i-j|)=\sum_{i,j}\langle(n_{i,xz}-n_{i,yz})(n_{j,xz}-n_{j,yz})\rangle$,
and the calculated results indicate no obvious long-range nematic order in the
studied system. Hence, we conclude that the enhanced nematic correlation mainly
comes from short-range nematic fluctuations.

Secondly we investigate the responses of magnetic order on the onset of
nematic fluctuation. For most FeSCs, the antiferromagnetic (AFM) orders
usually locate near nematic and superconducting regimes in the phase diagram.
Previous QMC studies on the two-orbital Hubbard models suggest a robust ($\pi$,0)
or (0,$\pi$) AFM order upon increasing the on-site Hubbard U~\cite{Liu2014}. As shown
in Fig.~\ref{struc_factor1}(b), the on-site nematic interaction g clearly suppresses
($\pi$,0) magnetic order, which is reasonable since the on-site nematic interaction in
Eq.~(\ref{hint2}) effectively reduces the strength of Hubbard U.

In Fig.~\ref{struc_factor2} we present the nematic and spin structure
factors by using Raghu hopping parameters, which give a good description for iron
pnictides. Figure ~\ref{struc_factor2}(a) demonstrates a similar but much more clear
($\pi$,$\pi$) nematic fluctuation upon increasing the on-site nematic interaction strength g.
Figure ~\ref{struc_factor2}(b) shows that the spin structure factor is
depressed by the on-site nematic interaction, especially for the ($\pi$,0) point.
Based on the results from Fig.~\ref{struc_factor1} and Fig.~\ref{struc_factor2}, we conclude
that the on-site nematic interaction acts to enhance nematic fluctuations and suppress AFM
spin fluctuations in FeSCs.

\begin{figure}
\centering
\includegraphics[scale=0.62]{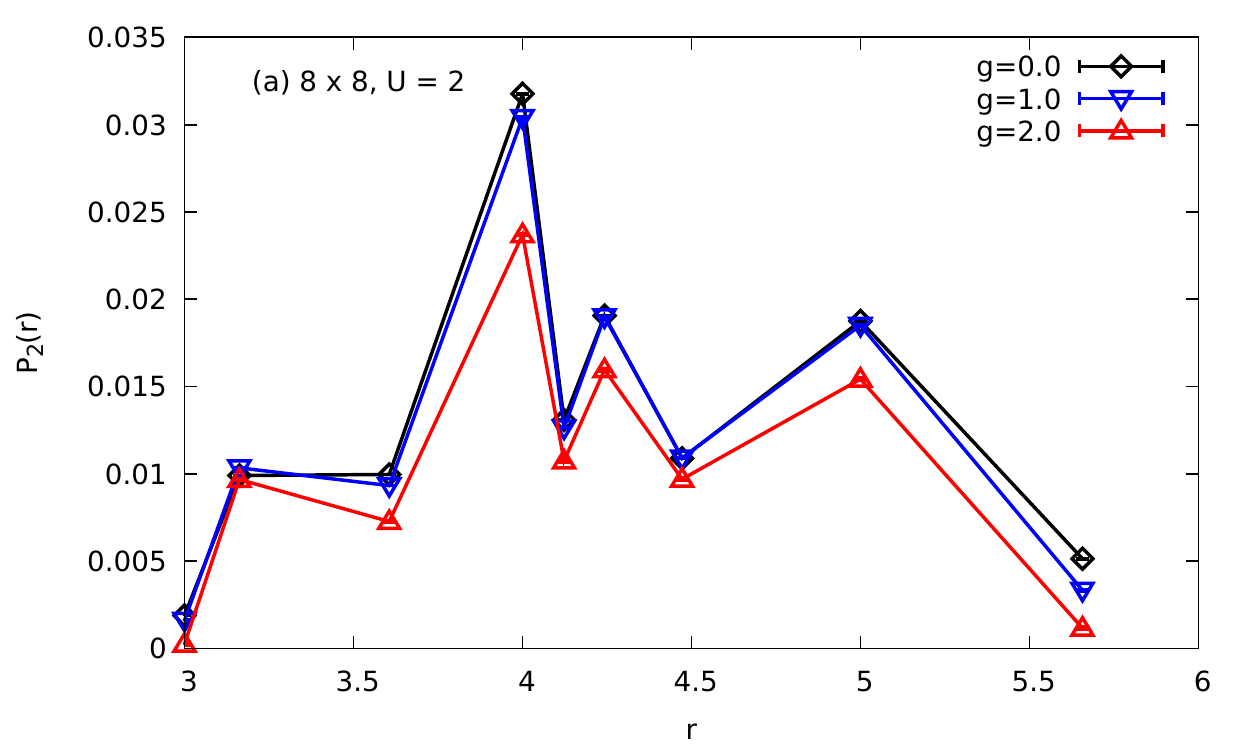}
\includegraphics[scale=0.62]{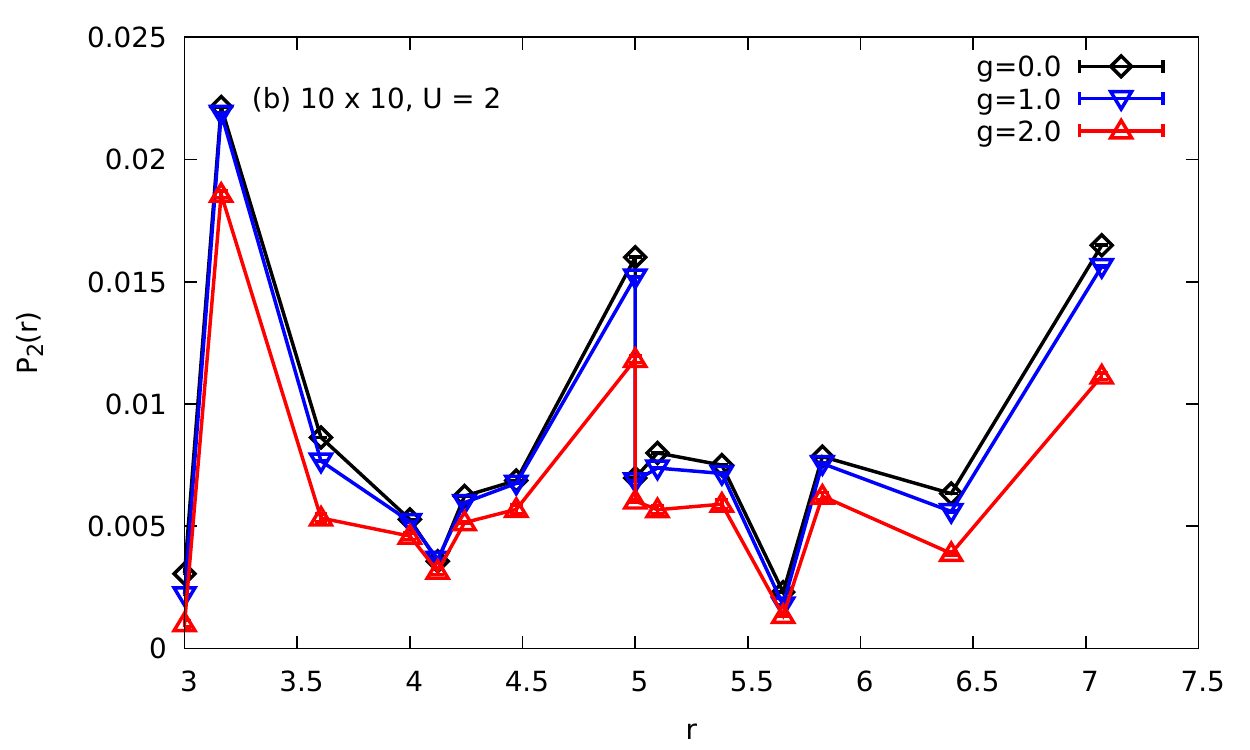}
\caption{(color online)
Long-range correlation function of wave2 versus pairing distance r by using Dumitrescu
hopping parameters.
(a) Hole doping density $\rho=0.125$ and U = 2 eV on an 8$\times$8 lattice;
(b) Hole doping density $\rho=0.08$ and U = 2 eV on a 10$\times$10
lattice.} \label{pairing1}
\end{figure}

\begin{figure}
\centering
\includegraphics[scale=0.62]{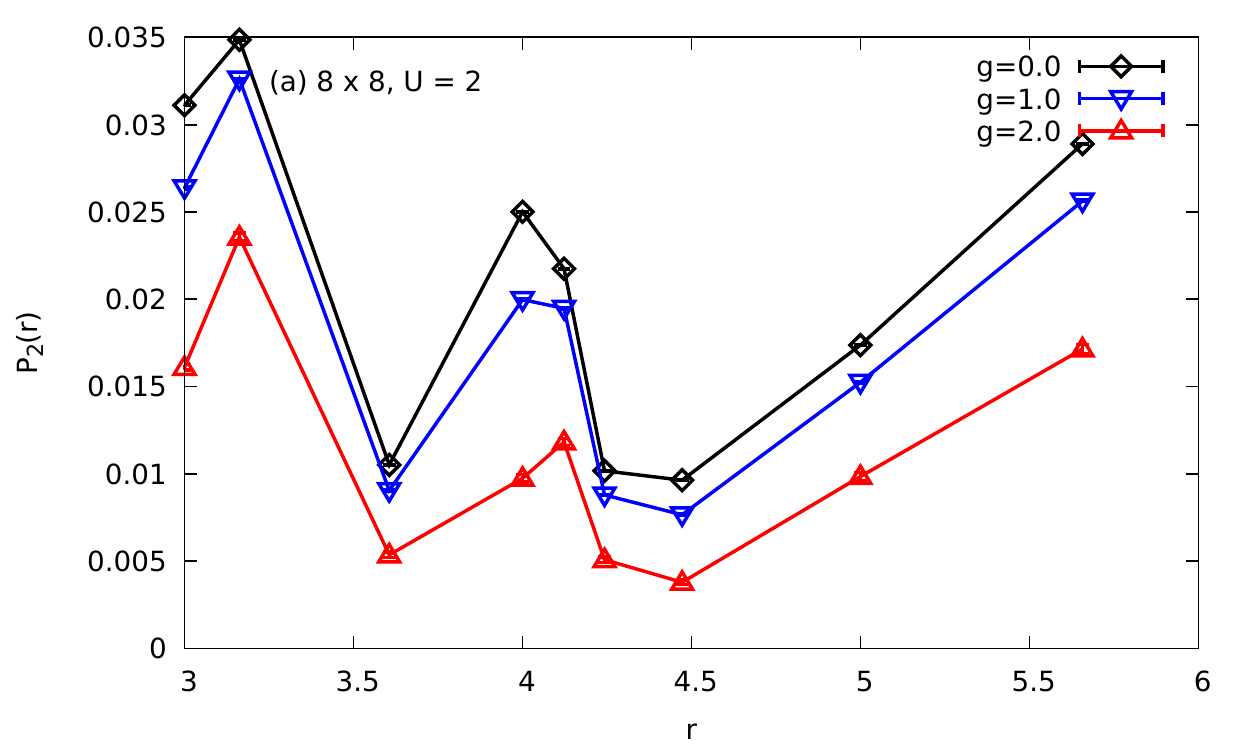}
\includegraphics[scale=0.62]{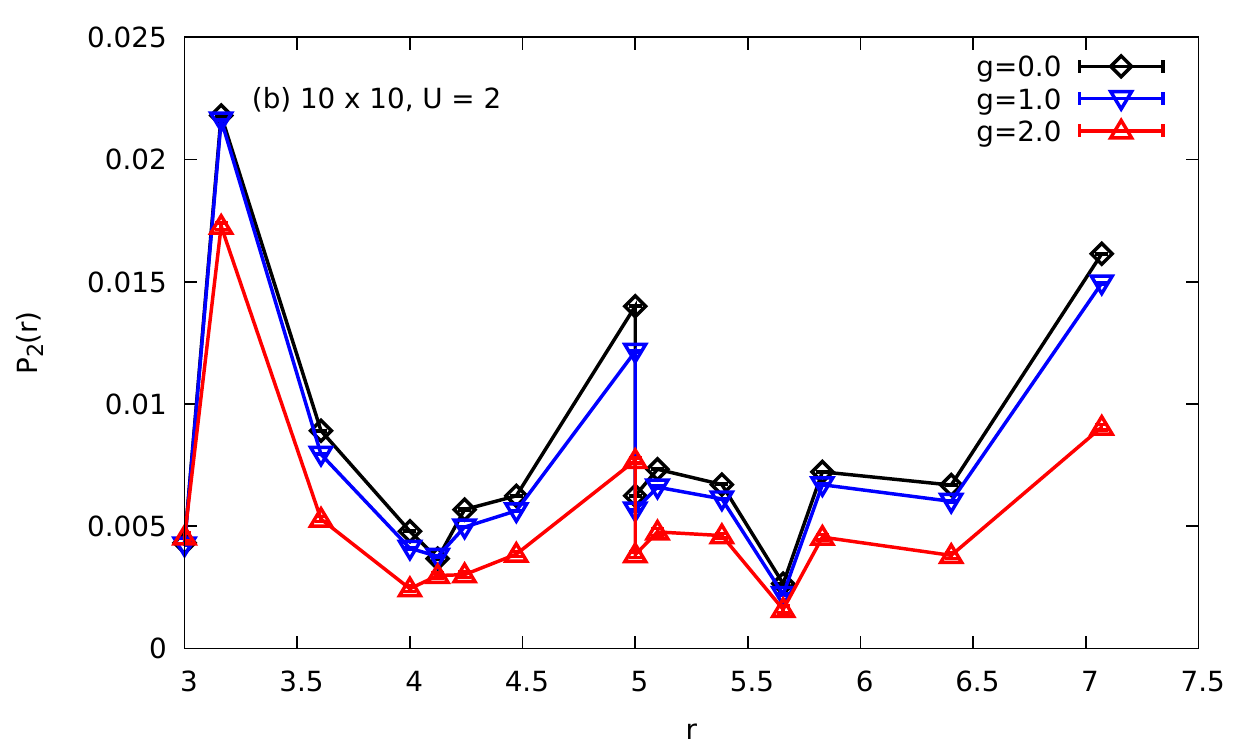}
\caption{(color online)
Long-range correlation function of wave2 versus pairing distance r by using Raghu
hopping parameters.
(a) Hole doping density $\rho=0.125$ and U = 2 eV on an 8$\times$8 lattice;
(b) Hole doping density $\rho=0.08$ and U = 2 eV on a 10$\times$10
lattice.} \label{pairing2}
\end{figure}

\subsection{Pairing correlations}
In this section, we will discuss another important issue about the influence of the
on-site nematic interaction on electron pairings. We first briefly discuss the pairing
operators for multi-orbital models. As
shown in Eq.~(5), the pairing operators in the two-orbital model consist of not only
the spatial- but also the orbital- distributions, in which the factor $f(\textbf{q})$
is the spatial part and $\tau_i$ stands for the orbital distribution. In general, there
will be dozens of pairing candidates in the two-orbital model.

In Fig.~\ref{ave_p1}(a), we show $P_{all}$ for four typical pairings, wave2, wave3, wave6,
and s$_{\pm}$, whose definitions could be reached in Table~I and Eq.~(5). Pairing wave2
with A$_{1g}$ symmetry is one of pairings with large pairing amplitude~\cite{Liu2014}.
Pairing s$_{\pm}$ also has large amplitude and is one of the most possible pairing candidates
in FeSCs. Pairing wave3 was studied in the simple two-orbital model~\cite{Dumitrescu2016}.
One can see that $P_{all}$ for wave2 and s$_{\pm}$ is decreased with increasing the on-site
nematic interaction g, whereas wave3 and wave6 exhibit an opposite behavior. Note that the
enhancement of $P_{all}$ for wave3 is in agreement with previous finding in the simple
two-orbital model~\cite{Dumitrescu2016}, in which Coulombic interactions were neglected.
We would like to point out all other pairings not shown here are suppressed by the on-site
nematic interaction.

Since the short-ranged pairing correlations contain the contributions from local spin and/or
charge components~~\cite{Huang2001a,Huang2001b}, in certain cases they may mislead our
understanding on the intrinsic superconducting property. To exclude the effect of short-ranged
pairing correlations, we show the long-distance averaged pairing correlation $P_{long}$ in
Fig.~\ref{ave_p1}(b). There are two significant differences compared with $P_{all}$ in
Fig.~\ref{ave_p1}(a): (1) The dominant pairing seems to be s$_{\pm}$-wave, instead of wave2;
(2) All pairing channels respond negatively to the on-site nematic interaction g. We found
that all the pairing channels, including other ones not presented here, are suppressed by
g. In particular, pairing wave3 and wave6 show different behaviors with increasing g for
all-distance and long-distance averages pairing correlations. These differences are induced by
short-ranged contribution of wave3 and wave6, which usually has much larger amplitude than
the long-distance counterpart.

We also calculated the all-distance and long-distance averaged pairing correlations by using
Raghu hopping parameters, and the obtained results are shown in Figs.~\ref{ave_p2}(a)
and (b). Similar suppression of $P_{long}$ by the on-site nematic interaction is clearly observed.
One difference from Dumitrescu hopping parameters is that wave2 with A$_{1g}$ symmetry seems to
be the dominant pairing channel from both all-distance and long-distance averaged pairing
correlations. In addition, $P_{all}$ for wave3 and wave6 is no longer enhanced by g.
The universal suppression of long-range pairing correlations by g suggests that the main effect
of the on-site nematic interaction is to suppress superconductivity in the studied model.

In order to clearly demonstrate the long-range pairing behavior on g, we pick the dominant pairing
wave2 as an example and investigate the pairing distance dependence of long-range pairing
correlation of wave2. Figs.~\ref{pairing1}(a) and (b) show the long-range pairing correlation
P$_{2}$(r) as a function of r by using Dumitrescu hopping parameters on the 8$\times$8 and
10$\times$10 lattices under various nematic interaction strengths, respectively. One can readily
see a suppression of P$_{2}$(r) at different distances as g is increased. Similar results by using
Raghu hopping parameters are shown in Figs.~\ref{pairing2} (a) and (b). Obviously, the on-site
nematic interaction g still acts to suppress P$_{2}$(r) at different distances.

Why are the long-range pairing correlations suppressed by the on-site nematic interaction g?
One possible reason is that the decrease of spin fluctuations around ($\pi$,0)/(0,$\pi$) leads to
strong reduction of pairing amplitude for several dominant pairing channels, which may overwhelm
the contribution to electron pairing from enhanced nematic fluctuations. Another possible reason is
that the enhanced nematic fluctuations suppress the phase coherence between electron pairs.

\section{Conclusions}
We studied the nematic, magnetic, and pairing properties of the two-orbital Hubbard model that
consists of Coulombic interactions and on-site nematic interaction. The main advantage of our model
is that we could completely study the impact of nematic interaction on electron pairings by taking
electronic correlations into account.

Our results based on the CPQMC simulations indicate that the on-site nematic interaction seems to
prompt antiferro-orbital nematic fluctuations and suppress the ($\pi$,0)/(0,$\pi$) AFM order.
Most importantly, the universal suppression of $P_{long}$ for several dominant pairing channels by g
suggests that the enhancement of nematic fluctuation plays a negative role on superconductivity.
Our finding is useful for understanding the interplay of nematic fluctuation and superconductivity
in FeSCs.

\section*{Acknowledgments}
G.L. thanks Yan Zhang and Yong-Jun Wang for insightful discussions. This work was
supported by the National Natural Science Foundation of China under grant No.~11674087.




\section*{References}
\begin{thebibliography}{}

\bibitem{Dagotto2013}
E. Dagotto, Rev. Mod. Phys. \textbf{85}, 849 (2013).

\bibitem{Johnston2010}
D. C. Johnston, Adv. Phys. \textbf{59}, 803 (2010).

\bibitem{Hirschfeld2011}
P. J. Hirschfeld, M. M. Korshunov, and I. I. Mazin, Rep.
Prog. Phys. \textbf{74}, 124508 (2011).

\bibitem{Dai2012}
P. Dai, J. P. Hu, and E. Dagotto, Nat. Phys. \textbf{8}, 709 (2012).

\bibitem{Lawler2010}
M. J. Lawler, K. Fujita, J. Lee, A. R. Schmidt, Y. Kohsaka,
C. K. Kim, H. Eisaki, S. Uchida, J. C. Davis, J. P. Sethna,  and
E.-A. Kim, Nature \textbf{466}, 347 (2010).

\bibitem{Fischer2011}
M. H. Fischer and E.-A. Kim, Phys. Rev. B \textbf{84}, 144502 (2011).

\bibitem{Edegger2006}
B. Edegger, V. N. Muthukumar,  and C. Gros, Phys. Rev. B \textbf{74},
165109 (2006).

\bibitem{Sun2010}
K. Sun, M. J. Lawler,  and E.-A. Kim, Phys. Rev. Lett. \textbf{104}, 1
(2010).

\bibitem{Zheng2014}
X.-J. Zheng, Z.-B. Huang,  and L.-J. Zou, J. Phys. Soc. Japan
\textbf{83}, 24705 (2014).

\bibitem{Zheng2015}
X.-J. Zheng, Z.-B. Huang,  and L.-J. Zou, Phys. Rev. B
\textbf{92}, 085109 (2015).

\bibitem{Chu2010}
J.-H. Chu, J. G. Analytis, K. De Greve, P. L. McMahon, Z. Is-
lam, Y. Yamamoto,  and I. R. Fisher, Science \textbf{329}, 824 (2010).

\bibitem{Dusza2011}
A. Dusza, A. Lucarelli, F. Pfuner, J.-H. Chu, I. R. Fisher, and
L. Degiorgi, Eur. Lett. \textbf{93}, 37002 (2011).

\bibitem{Fernandes2014}
R. M. Fernandes, A. V. Chubukov, and J. Schmalian,
Nature Physics \textbf{10}, 97 (2014).

\bibitem{Kaczmarczyk2016}
J. Kaczmarczyk, T. Schickling, and J. B\"unemannu,
Phys. Rev. B \textbf{94}, 085152 (2016)

\bibitem{McQueen2009}
T.M. McQueen, A.J. Williams, P.W. Stephens, J. Tao, Y. Zhu, V. Ksenofontov,
F. Casper, C. Felser, and R.J. Cava, Phys. Rev. Lett. \textbf{103}, 057002 (2009)

\bibitem{Kim2014}
H. Kim, M. A. Tanatar, W. E. Straszheim, K. Cho, J. Murphy, N. Spyrison,
J.-Ph. Reid, B. Shen, H.-H. Wen, R. M. Fernandes, and R. Prozorov,
Phys. Rev. B \textbf{90}, 014517 (2014)

\bibitem{Cai2014}
P. Cai, W. Ruan, X. Zhou, C. Ye, A. Wang, X. Chen, D.-H. Lee, Y. Wang,
Phys. Rev. Lett. \textbf{112} 127001 (2014).

\bibitem{Moon2012}
E.-G. Moon and S. Sachdev,
Phys. Rev. B \textbf{85}, 184511 (2012).


\bibitem{Yuan2016}
D. Yuan, J. Yuan, Y. Huang, S. Ni, Z. Feng, H. Zhou, Y. Mao, K. Jin, G. Zhang,
X. Dong, F. Zhou, and Z. Zhao, Phys. Rev. B \textbf{94}, 060506(R) (2016).

\bibitem{Lederer2015}
S. Lederer, Y. Schattner, E. Berg, and S.A. Kilvelson,
Phys. Rev. Lett. \textbf{114}, 097001 (2015).

\bibitem{Fernandes2011}
R. M. Fernandes, E. Abrahams, and J. Schmalian, Phys. Rev.
Lett. \textbf{107}, 217002 (2011).

\bibitem{Fernandes2012}
R. M. Fernandes, A. V. Chubukov, J. Knolle, I. Eremin, and J.
Schmalian, Phys. Rev. B \textbf{85}, 024534 (2012).

\bibitem{Bishop2016}
C.B. Bishop, A. Moreo, and E. Dagotto,
Phys. Rev. Lett. \textbf{117},117201 (2016).

\bibitem{Bishop2017}
C.B. Bishop, J. Herbrych, E. Dagotto and A. Moreo,
Phys. Rev. B \textbf{96},035114 (2017).

\bibitem{Chubukov2012}
A. Chubukov, Ann. Rev. Conden. Matt. Phys.
\textbf{3},57 2012.

\bibitem{Lee2012}
W.-C. Lee and P. W. Phillips, Phys. Rev. B \textbf{86}, 245113 (2012).

\bibitem{Lv2009}
W Lv, J Wu, P Phillips,
Phys. Rev. B \textbf{80} 224506 (2009)

\bibitem{Lv2010}
W Lv, F Kr\'uger, P Phillips,
Phys. Rev. B \textbf{82} 045125 (2010)

\bibitem{Yamase2013}
H. Yamase, and R. Zeyher, Phys. Rev. B \textbf{88}, 180502(R) (2013).

\bibitem{Dumitrescu2016}
Philipp T. Dumitrescu, Maksym Serbyn, Richard T. Scalettar, and Ashvin Vishwanath,
Phys. Rev. B \textbf{94} 155127 (2016)

\bibitem{Raghu2008}
S. Raghu, Xiao-Liang Qi, Chao-Xing Liu, D. J. Scalapino, and Shou-Cheng Zhang,
Phys. Rev. B \textbf{77} 220503(R) (2008)

\bibitem{Liu2016}
Guangkun Liu, Nitin Kaushal, Shaozhi Li, Christopher B. Bishop, Yan Wang,
Steve Johnston, Gonzalo Alvarez, Adriana Moreo, and Elbio Dagotto,
Phys. Rev. E \textbf{93} 063313 (2016)

\bibitem{Liu2014}
Guangkun Liu, Zhongbing Huang, Yongjun Wang, J. Phys.: Condens. Matter,
\textbf{26} 325601 (2014).

\bibitem{Daghofer2008}
M. Daghofer, A. Moreo, J. A. Riera, E. Arrigoni, D. J. Scalapino, and E. Dagotto,
 Phys. Rev. Lett. \textbf{101} 237004 (2008)

\bibitem{Luo2010}
Qinlong Luo, George Martins, Dao-Xin Yao, Maria Daghofer, Rong Yu, Adriana Moreo, Elbio Dagotto,
Phys. Rev. B \textbf{82} 104508 (2010)

\bibitem{Wan2009}
Y. Wan and Q.-H. Wang, Europhys. Lett. \textbf{85} 57007 (2009)

\bibitem{Moreo2009}
Adriana Moreo, Maria Daghofer, Andrew Nicholson, and Elbio Dagotto,
 Phys. Rev. B \textbf{80} 104507 (2009).

\bibitem{Zhang1997a}
Shiwei Zhang, J. Carlson, and J. E. Gubernatis, Phys. Rev. B \textbf{55},
7464 (1997).

\bibitem{Zhang1997b}
Shiwei Zhang, J. Carlson, and J. E. Gubernatis, Phys. Rev. Lett.
\textbf{78}, 4486 (1997).

\bibitem{Huang2001a}
Z.B. Huang, H.Q. Lin, J.E. Gubernatis, Phys. Rev. B \textbf{63}, 115112 (2001)

\bibitem{Huang2001b}
Z.B. Huang, H.Q. Lin, J.E. Gubernatis, Phys. Rev. B \textbf{64}, 205101 (2001)

\bibitem{Sakai2004}
S. Sakai, R. Arita and H. Aoki, Phys. Rev. B \textbf{70}, 172504  (2004)

\end{thebibliography}
\end{document}